\documentclass[12pt]{iopart}

\usepackage{graphicx}

\begin{document}

\title{Theoretical calculations for solid oxygen under high pressure}

\author{Kazuki Nozawa, Nobuyuki Shima and Kenji Makoshi}

\address
{
Graduate School of Material Science, University of Hyogo, 3-2-1, Kouto,
 Kamigohri, Hyogo 678-1297, Japan \\
}

\date{\today}

\begin{abstract}
The crystal structure of solid oxygen at low temperatures and at pressures up to 7 GPa
 is studied by theoretical calculations.
In the calculations, the adiabatic potential of the crystal is approximated by
 a superposition of
 pair-potentials between oxygen molecules calculated by an {\it ab-initio} method.
The monoclinic $\alpha$ structure is stable up to 6 GPa and calculated lattice parameters
 agree well with experiments.
The origin of a distortion and that of 
 an anisotropic lattice compressibility
 of the basal plane of $\alpha$-O$_2$ are clearly demonstrated.
In the pressure range from 6 to 7 GPa, two kinds of structures are proposed by
X-ray diffraction experiments: the $\alpha$ and orthorhombic $\delta$ structures.
It is found that the energy difference between
 these structures becomes very small in this pressure range.
The relation between this trend and the incompatible results of X-ray diffraction
experiments is discussed.

\end{abstract}


\maketitle

\section{Introduction}
At low temperatures or under pressures, molecular oxygen is solidified 
 by weak intermolecular interactions.
At zero pressure, oxygen transforms to monoclinic $\alpha$-O$_2$ through
 $\gamma$- and $\beta$-O$_2$ by cooling.
In the $\alpha$ phase, oxygen molecules condense with its molecular axis
 perpendicular to the basal plane of the monoclinic lattice.
As shown in figure~1(a), experiments demonstrated that the unit cell 
includes two oxygen molecules
 and the structure belongs to $C2/m$ space group~\cite{barrett}.
The ground electronic state of the oxygen molecule is spin triplet state.
The molecular spins, which are perpendicular to the molecular axis,
 order antiferromagnetically on the basal plane of $\alpha$-O$_2$ with 
 easy axis parallel to the $b$-axis.
The basal plane is illustrated in figure~1(b). Arrows in the figure
 show the directions of magnetic moments.
The crystal structures of solid oxygen might be 
 correlated with the magnetic moments even at high pressures.

A Raman experiment reported a phase transition from 
 $\alpha$-O$_2$ to another monoclinic or
 orthorhombic structures below 3 GPa~\cite{jodl}.
X-ray diffraction experiments using high-brilliance
 synchrotron radiation source, however, reported different phase diagrams.
Akahama {\it et al.} reported $\alpha$-O$_2$ transforms to $\epsilon$-O$_2$ directly
 at 7.2 GPa at 19 K~\cite{akahama}.
Although they also found an anomaly in lattice constants just before the transition
 to $\epsilon$-O$_2$, it was not considered as a sign of a transition to other phases.
Gorelli {\it et al.} also reported the stability of $\alpha$-O$_2$
 up to about 5.3 GPa, and
they found orthorhombic $\delta$-O$_2$ (space group: {\it Fmmm})
 above 5.3 GPa at 65 K~\cite{gorelli}.
Although the monoclinic $\alpha$ structure is ascertained to be stable up to about 5 GPa
 at low temperatures in both experiments,
 the stable structure at pressures between 5-7 GPa is considered
 to be still open question. 
For higher pressures, recent X-ray diffraction experiments have revealed that 
$\epsilon$-O$_2$ consists  of O$_8$ clusters~\cite{fujihisao8, natureo8}.
The $\epsilon$ phase transforms to the metallic $\zeta$ phase at higher pressure
($\sim$ 100 GPa) and finally to
 a superconducting state~\cite{desgreniers,akahamaz,shimizu}.
The mechanism of the transition from or to the $\epsilon$ phase and the structure
of the $\zeta$ phase are still unknown.

Regarding theoretical studies, Etters {\it et al.} obtained
 the structure of $\alpha$-O$_2$ 
 using semi-empirical pair-potentials including magnetic interactions
 between O$_2$ pairs and predicted a phase transition from the $\alpha$ to
 an orthorhombic phase at 2.3 GPa~\cite{ettersdelta}.
We performed theoretical calculations using {\it ab-initio} pair-potentials and 
reported that the monoclinic $\alpha$ structure is stable up to 6 GPa~\cite{nozawa}.
A part of the results will be presented in this paper.
First-principles investigations based on density functional theory 
(DFT)~\cite{dft} were performed especially for higher pressure phases 
$\epsilon$ and $\zeta$~\cite{serra,otani,kususe,gebauer,neaton,ma}. 
The predicted structures for insulating $\epsilon$-O$_2$ is however 
not consistent with the experimental one. This is caused by a failure of 
the local density approximation (LDA) in describing the magnetic 
interaction between oxygen molecules~\cite{thnoza}. Because the failure does 
not influence the metallic state calculations, the structural relaxation 
based on DFT with an appropriate starting structure can predict the 
structure of the $\zeta$ phase. A recent first-principles study reported a 
plausible structure for the metallic $\zeta$ phase~\cite{ma}. In this paper, 
we study the insulating $\alpha$ and $\delta$ phases. The total energy is 
evaluated as a superposition of pair-potentials obtained by quantum 
chemistry calculations including configuration interactions.

\section{Theory and Calculation}
Under low pressures, solid oxygen consists of weakly bound oxygen molecules.
Therefore the interactions between molecules need to be reproduced exactly
in order to calculate the structure.
The total energy calculation based on DFT is often used to
 investigate structures of various crystals.
Unfortunately, DFT calculations within the LDA (or with gradient corrections) does not
 give correct intermolecular interactions for magnetic oxygen molecules.
It is related with the symmetries of the highest occupied molecular
orbital (HOMO) and lowest unoccupied molecular orbital (LUMO), and the well-known 
bandgap-underestimation problem of the LDA~\cite{thnoza}.
Both of the HOMO and LUMO of interacting oxygen molecules 
are originated in the $\pi_g$ (gerade) orbital of O$_2$. Thus the 
underestimation of the HOMO-LUMO gap cause a serious overestimation 
of the exchange energy.
We actually confirm DFT calculations do not reproduce the $\alpha$ structure at
zero pressure.
In many cases, quantum chemistry calculations including the configuration interactions
reproduce intermolecular interactions correctly.
In this paper, we evaluate the total energy as a superposition of intermolecular
potentials (pair-potentials) calculated by an {\it ab-initio} method with 
configuration interactions.

In order to reduce degrees of freedom of the O$_4$ system,
we introduce following assumptions into the calculations.
\begin{enumerate}
\item An antiferromagnetic ordering in the $ab$ (basal) plane.
\item The $C2/m$ lattice symmetry.
\item A molecular axis perpendicular to the $ab$ plane.
\end{enumerate}
These assumptions are consistent with known properties of the structure of $\alpha$-O$_2$,
and we can treat the orthorhombic $\delta$ structure as a special case that the monoclinic
angle $\beta^* = 90 ^\circ$.

In the orthorhombic structure, two types of the magnetic ordering along the $c$-axis
are allowed: the ferromagnetic or anti-ferromagnetic orderings.
In the $\alpha$ phase, the magnetic ordering along the $c$-axis is obviously
 ferromagnetic because of the periodicity along $c$-direction.
Although recent neutron diffraction experiment reported anti-ferromagnetic ordering
along the $c$-axis in the $\delta$ phase~\cite{goncharenkodelta}, 
we assume the ferromagnetic ordering along the $c$-axis 
because it simplify the treatment of the transition
from $\alpha$ to $\delta$ phases.
We have however confirmed that the difference in the total energy between the
 two kinds of orderings on the same lattice is very small.
It is ascribed to the weak magnetic interaction between molecules in different $c$-layers.

Following the assumption 1, there are two kinds of pairs of oxygen molecules.
One is the pair in which magnetic moments are parallel
 (ferromagnetic pair, F).
The other is the pair in which magnetic moments are anti-parallel
 (antiferromagnetic, AF).
Assumptions 2 and 3 restrict geometric configuration of molecules.
In addition to these assumptions, we fix the inter-atomic distance in O$_2$.
By these constraints, degrees of freedom of the O$_4$ system, in other words the number
 of geometric parameters of pair-potentials, are reduced from twelve to two.
Remaining degrees of freedom are the relative position of molecules.
The relative position of oxygen molecules ($x, z$) is defined as given in figure~2.
Consequently, the total energy is written as
\begin{equation}
 U_{\rm total} = \sum_{l, m, n} [ U^{\rm F}( x^{\rm F}_{lmn}, z_n )
 + U^{\rm AF}(x^{\rm AF}_{lmn}, z_n )] ~.
   \label{etot} 
\end{equation}
In the equation, pair-potentials $U^{\rm F}$ and $U^{\rm AF}$ denote
 interactions of the F and AF pairs, respectively.
$l,m$ and $n$ are indices of lattice vectors and coordinates
 $x^{\rm F}_{lmn}, x^{\rm AF}_{lmn}$ and $z_n$ are given as
\begin{eqnarray}
x_{lmn}^{\rm F}&=& \sqrt{\left(la -nc\cos\beta\right)^2
+\left(mb\right)^2}~,\\
x_{lmn}^{\rm AF} &=& \sqrt{\left(\left(l+\frac{1}{2}\right)a
 -nc\cos\beta\right)^2 +\left(\left(m+\frac{1}{2}\right)
 b\right)^2}~,\\
z_n &=& |nc\sin\beta|~.
\end{eqnarray} 
Here, $a, b, c$ and $\beta$ are lattice constants and the monoclinic angle, respectively. 
Pair-potentials $U^{\rm F}$ and $U^{\rm AF}$ are calculated by a complete active space
 self-consistent field (CASSCF) method~\cite{casscf}.
The CASSCF calculation is performed with GAMESS program~\cite{gamess} using
 3s2p1d atomic natural orbital basis set~\cite{ano}.
Details of calculations and obtained pair-potentials have been 
 given in our previous brief report~\cite{nozawa}.
\section{Results and Discussions}
\subsection{Structures up to 6 GPa and its origin}
A Raman experiment reported $\alpha$-O$_2$ transforms to another monoclinic
 structure or orthorhombic structure under 3 GPa~\cite{jodl}.
X-ray diffraction experiments by Akahama {\it et al.}
 and Gorelli {\it et al.}, however, showed the $\alpha$ structure is stable
 up to higher pressure.
In this section, we present a result of the theoretical calculations that
the $\alpha$ structure is stable up to 6 GPa.

The calculated pressure dependences of lattice parameters up to 6 GPa are
 shown in figures~3(a),(b) and (c).
The results of higher pressure, which is also shown in the figures,
 will be discussed later.
In figures, solid lines present calculated lattice parameters, and
 closed circles, triangles and diamonds are experimental results~\cite{barrett,akahama}.
The values of pressure are evaluated by the 
numerical differentiation of the total energy with respect to the volume 
($P = - {\partial E}/{\partial V}$).

As previously mentioned,
 Raman and theoretical studies reported phase transitions in this pressure 
range~\cite{jodl,ettersdelta}.
The structural transition predicted by the theoretical calculation was accompanied 
by abrupt changes of lattice parameters at 2.3 GPa. 
As shown in figure~3, present results decrease continuously
 with increasing pressure and there are no
 sign of the structural transition to the $\delta$ structure up to 6 GPa.

As shown in figure~3(a), the crystal is less compressible along the $b$-axis than
 the $a$-axis.
The relation of the anisotropic lattice compressibility and the AF ordering
 in the $ab$ plane is discussed by Akahama {\it et al}~\cite{akahama}.
The $b/a$ ratio is 0.638 at zero pressure,
 and it increases with pressure and reaches 0.691 at 6 GPa.
The experimental value is 0.635 at zero 
 pressure~\cite{barrett}, and it reaches 0.687 at 6 GPa~\cite{akahama}.
The agreement between the calculation and experiments is very satisfactory.
As previously reported, the crystal structure of $\alpha$-O$_2$
 is considered as a result of magnetic interactions between
 oxygen molecules~\cite{ettersdelta}.
In order to clarify how the magnetic interactions affect the crystal structure,
 we consider a simple model of the $ab$ plane as shown in figure~4.
This figure shows the top view of the $ab$ plane. The molecular axes are 
perpendicular to the figure, and
 arrows represent magnetic moment directions of O$_2$ molecules.
The unit cell contains ten pair of oxygen molecules: four AF pairs at a 
distance $r=\sqrt{a^2+b^2}/2$ and three kinds of F pairs at distances 
$a$, $b$ and $2r$. 
For simplicity, we neglect interactions between molecules in which
 intermolecular distances are $2r$.
Consequently the (total) energy of this system is written as
\begin{eqnarray}
\label{etotmodel}
E(r, b) &=& 4A(r)+2\Bigl[F(\sqrt{4r^2-b^2})+F(b)\Bigr] ~,
\end{eqnarray}
where $A(x)$ and $F(x)$ denote interactions (pair-potentials) of the AF and F pairs
 at a distance $x$.
Calculated intermolecular potentials $A(x)$ and $F(x)$,
 which correspond to $U^{\rm AF}(x, 0)$ and $U^{\rm F}(x, 0)$ in (\ref{etot}),
 are shown in figure~5.
Optimal energy of (\ref{etotmodel}) are obtained at
 ($a, b, r$) = (5.42, 3.41, 3.20), and those are indicated by arrows in the figure.
The obtained values are very close to the experimental values at
 zero pressure (5.40, 3.43, 3.20)~\cite{barrett}.

Evidently (\ref{etotmodel}) is dominated by the first term because of its coefficient
and the fact that $A(x)$ has deeper minimum than $F(x)$.
Thus we obtain the optimal value of $r$, $r_0 \equiv 3.2$ where
 $A(r)$ takes the lowest energy.
Then the problem is simplified as $E(r_0, b) = 4A(r_0)+2E_2(b)$, where
$E_2(b) = F(a(b))+F(b)$ and $a(b)=\sqrt{4r_0^2-b^2}$.
$E(r_0, b)$ takes a minimum on the condition $\frac{{\rm d}E_2(b)}{{\rm d}b}=0$, i.e.,
\begin{equation}
\label{cond}
a(b) \frac{{\rm d}F(b)}{{\rm d}b} = b \frac{{\rm d}F(a)}{{\rm d}a} ~.
\end{equation}
This condition is symmetric for $b$ and $a$, then 
$a=b (=\sqrt{2}r_0)$ satisfies the condition. It gives, however, the highest
energy because $\frac{{\rm d}^2 F}{{\rm d}b^2}$ is negative at $b=\sqrt{2}r_0 \sim 4.5$
as shown in figure~5.
Another solution $b_0$ of (\ref{cond})
is located at slightly larger side of the minimum of $F(x)$: around 3.4. 
As a result, the optimal value of $r$ and $b$, the structure of the $ab$ plane therefore,
strongly depend on the position of the minimum of $A(x)$ and $F(x)$.
The difference of $A(x)$ and $F(x)$ corresponds to the exchange energy.
If there are no magnetic interactions between the molecules,
$A(x)$ is equivalent to $F(x)$.
In this case, following the above discussion, the optimal value of $r$ is equal
 to $b_0$ and it gives the close-packed triangular lattice.
The distortion is not introduced if there are only one kind of pair-potential.
Namely the origin of the distorted triangular lattice in the $\alpha$ phase
is attributed to the existence of the two kinds of pair-potentials,
 in other words, the magnetic interaction.

The anisotropic lattice compressibility
 of the $ab$ plane, which is pointed out by Akahama
 {\it et al.}~\cite{akahama}, can be easily understood from figure~5.
First we consider a situation that the lattice is compressed along the $a$-axis
 and the length along the $b$-axis is fixed.
In this case, $r$ decrease following the
relation $r=\sqrt{a^2+b^2}$ and the first term of (\ref{etotmodel}) increases 
by the repulsive interaction. 
Some part of the increased energy is however cancelled out
by the term of $F(a)$ because it is in the attractive region.
On the other hand, when the lattice is compressed along the $b$-axis and $a$-direction
is fixed,
 both term of $A(r)$ and $F(b)$ increase, and as a result the total energy
 $E$ increase steeply compared with the previous case.
This is the origin of the anisotropic lattice compressibility of the $\alpha$ phase.

\subsection{Structures above 6 GPa}
\label{sec6GPa}
X-ray diffraction experiments by Akahama {\it et al.} and Gorelli {\it et al.}
 showed the monoclinic $\alpha$ structure is stable up to
 higher pressure than the previously proposed transition pressure.
Around 6 GPa, however, they proposed different structures.
Akahama {\it et al.} reported a direct structural transformation from $\alpha$-
 to $\epsilon$-O$_2$ at 7.2 GPa.
On the contrary, Gorelli {\it et al.} found $\delta$-O$_2$
 between $\alpha$- and $\epsilon$-O$_2$.
In this section, we discuss the stability of $\alpha$- and $\delta$-O$_2$.

In figure~3, dashed lines above 6 GPa show lattice parameters of optimal
 structure. 
The dotted lines present extrapolated values from the data
 below 6 GPa, namely those form monoclinic $\alpha$ structure.
Although optimal values (dashed lines) show the orthorhombic structure
 is stable above 6 GPa, the energy of the orthorhombic structure is
 very close to that of the monoclinic (dotted lines) structure.

Since the $\delta$-O$_2$ can be considered as a special case of $\alpha$-O$_2$ 
($\beta^*$=90$^\circ$), we illustrate the total energy as a function of $\beta^*$
and pressure.
Figure~6 presents a pressure dependence of the cross section
 of the total energy from 4 GPa to 7 GPa.
In the figure, lattice constants $a$, $b$ and the volume of the unit cell are
 fixed to optimal values presented by the solid and dashed lines in figure~3(a).
Therefore the independent parameter is only $\beta^*$ at each pressure.
At the low pressure range, the total energy takes minima at two values of $\beta^*$,
 which give equivalent monoclinic $\alpha$ structures, and
 the orthorhombic structure ($\beta^*=90^\circ$) is unstable.
With adopted constraints of the crystal structure and geometric configurations of molecules,
 it can be derived analytically that
 a first derivative of the total energy with respect to $\beta^*$ is
 zero at $\beta^*=90^\circ$ (orthorhombic structure) at each pressure,
 i.e., the total energy takes a maximum or minimum values at the $\delta$
 structure~\cite{thnoza}.
With increasing pressure, the $\delta$ structure becomes stable at around
6 GPa in place of the $\alpha$ structure, the energy difference between 
these structure is however less than 0.5 meV ($\sim$ 5 K) even at 7 GPa.
Unfortunately, in the precision of present calculations,
 the energy difference may be too small to decide which structure is stable.
We obtained however an important result that the monoclinic-angle-dependence
 of the total energy is very small in this pressure range.
Akahama {\it et al.} and Gorelli {\it et al.} proposed different structures
 in this pressure range.
It is reasonable because the experiments were performed at temperatures higher than 
the energy difference we obtained.
The very small energy difference in this pressure range implies that
 several structures
 which have various monoclinic angles can be allowed at finite temperature.
Mita {\it et al.} performed Raman scattering experiments
 and observed peaks above 5 GPa~\cite{mita}.
Although the origin of the peaks is unresolved,
they indicated the existence of the complicated mixed phase at this pressure range.
Further investigations are desired to determine the origin of the unknown peaks.

At high temperatures, many experiments demonstrated the stability of
 orthorhombic $\delta$-O$_2$.
The stability of $\delta$-O$_2$ at high temperature is understood from figure~6.
As shown in the figure, the energy surface forms a double-well or
 parabolic shape.
In either case, they are symmetric
with respect to $\beta^*$ around $\beta^*=90^\circ$ and the energy barrier
separating the wells are very low.
Therefore the expectation value of $\beta^*$ becomes 90$^\circ$ even at low pressure
because of the effect of thermal excitations.

Akahama {\it et al.} reported abrupt changes of lattice
 parameters at the phase boundary of the $\alpha$ and $\epsilon$ phases.
In the present calculations, however, corresponding change of lattice parameters
are not obtained up to 12 GPa.
This may be caused by constraints on the crystal and/or magnetic structure
 in our calculations.
Recent neutron diffraction experiment shows a disappearance of the magnetic long-range
 ordering in the $\epsilon$ phase, but it does not exclude
the short-range order~\cite{goncharenkoeps}.
Intermolecular potentials adaptable to other magnetic configurations may be required 
to obtain the $\epsilon$ structure.

\section{Summary}
We investigated crystal structures of solid oxygen under pressure
 with {\it ab-initio} pair-potentials taking account of the difference of spin states.
Under 6 GPa, the obtained structure is monoclinic $\alpha$
 and calculated lattice parameters agree well with the results of the
 X-ray diffraction experiment.
The origin of the distortion of the basal plane and that of the anisotropic
 lattice compressibility of $\alpha$-O$_2$
 are clearly demonstrated that they are caused by the existence
 of two kinds of intermolecular interaction, namely magnetic interactions.
The crystal structure under the pressure range from 6 to 7 GPa is discussed using
 the total energy surface projected on the $P$-$\beta^*$ space.
Although orthorhombic $\delta$-O$_2$ is obtained as the stable structure, the energy 
difference between $\alpha$- and $\delta$-O$_2$ is less than 5 K in this pressure range.
This implies a possibility of a mixed phase including some structures which have
 various monoclinic angles.
 It seems to be consistent with recent X-ray diffraction and Raman experiments.
\section{Acknowledgements}
We are grateful to Y. Akahama for stimulating discussions and for providing
 us with experimental data.
We would like to thank Y. Ishii, H. Koizumi and Y. Mita for useful discussions and suggestions.
Numerical calculations were partially carried out using the computer facilities at the
 Research Center for Computational Science, Okazaki National Research Institutes.

\bigskip
\newpage
\noindent
{\bf FIGURE CAPTIONS}
\bigskip
\begin{description}

\item[figure~1]~~~
 (a) Monoclinic unit cell of $\alpha$-O$_2$.
 The unit cell includes two oxygen molecules.
 At $\beta^* = 90^\circ$, the unit cell becomes an
 orthorhombic $Fmmm$ structure.
 (b) The basal plane of $\alpha$-O$_2$. 
 Oxygen molecules are denoted by circles and 
 molecular axes are perpendicular to the $ab$ plane.
 Arrows represent magnetic moment directions. 

\item[figure~2]~~~
 A schematic view of coordinates for pair-potentials.
 The vector $(x, z)$ is the relative position of molecules.

\item[figure~3]~~~
 Pressure dependences of (a) lattice constants and monoclinic angles (b) $\beta$
 and (c) $\beta^*$ up to 7 GPa.
 Calculated values are denoted by solid lines and experimental values are represented
 by closed circles, triangles and diamonds.
 See text regarding the dashed and dotted lines.

\item[figure~4]~~~
 A model of the $ab$ plane.
 Circles denote oxygen molecules, and arrows represent magnetic moment directions.
 The unit cell includes four AF pairs and six F pairs.
 Two F pairs in which intermolecular distances are 2$r$ is not taken into account
 in the discussion.

\item[figure~5]~~~
 The optimized values of ($a, b, r$) and pair-potentials.
 These pair-potentials correspond to $U^{\rm F}(x, 0)$ and $U^{\rm AF}(x, 0)$, which are
 defined in previous section.
 Arrows denote the optimal values of ($a, b, r$).
 Note that the values of $b$ and $r$ are at almost minimum of pair-potentials.

\item[figure~6]~~~
 The contour plot of the total energy as a function of the pressure and $\beta^*$
 in the pressure range 4-7 GPa.
 The volume of the unit cell and lattice constants $a$ and $b$ are fixed at optimal
 values at each pressure, which are denoted by solid and dashed lines in figure~3.
 The difference in the energy between the monoclinic $\alpha$ and orthorhombic
 $\delta$ structures is very small at around 6 GPa.
 Contours increase by 2 meV.

\end{description}
\begin{figure}[p]
\includegraphics[width=8cm]{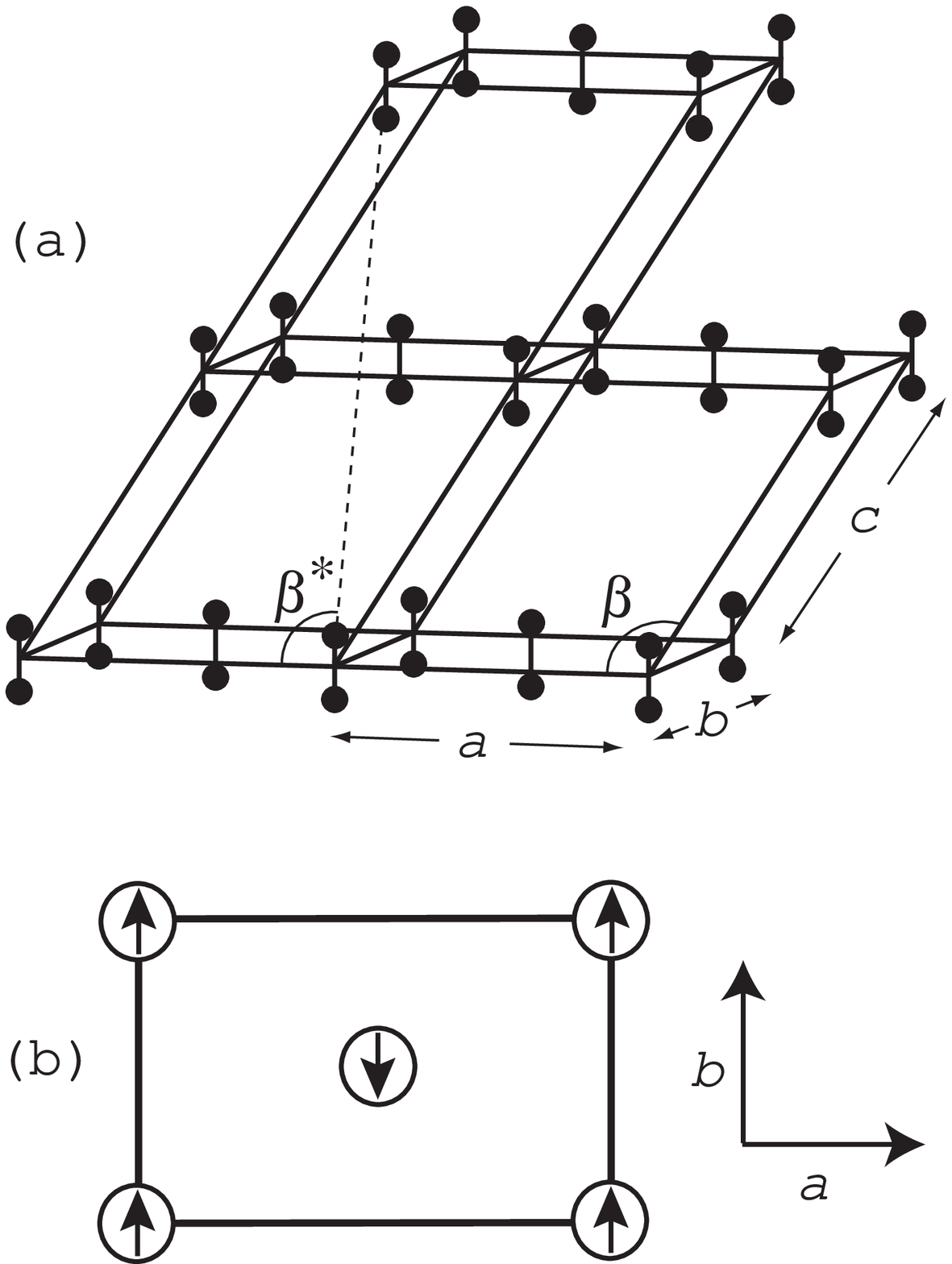}
\caption{}
\end{figure}
\begin{figure}[p]
\includegraphics[width=5cm]{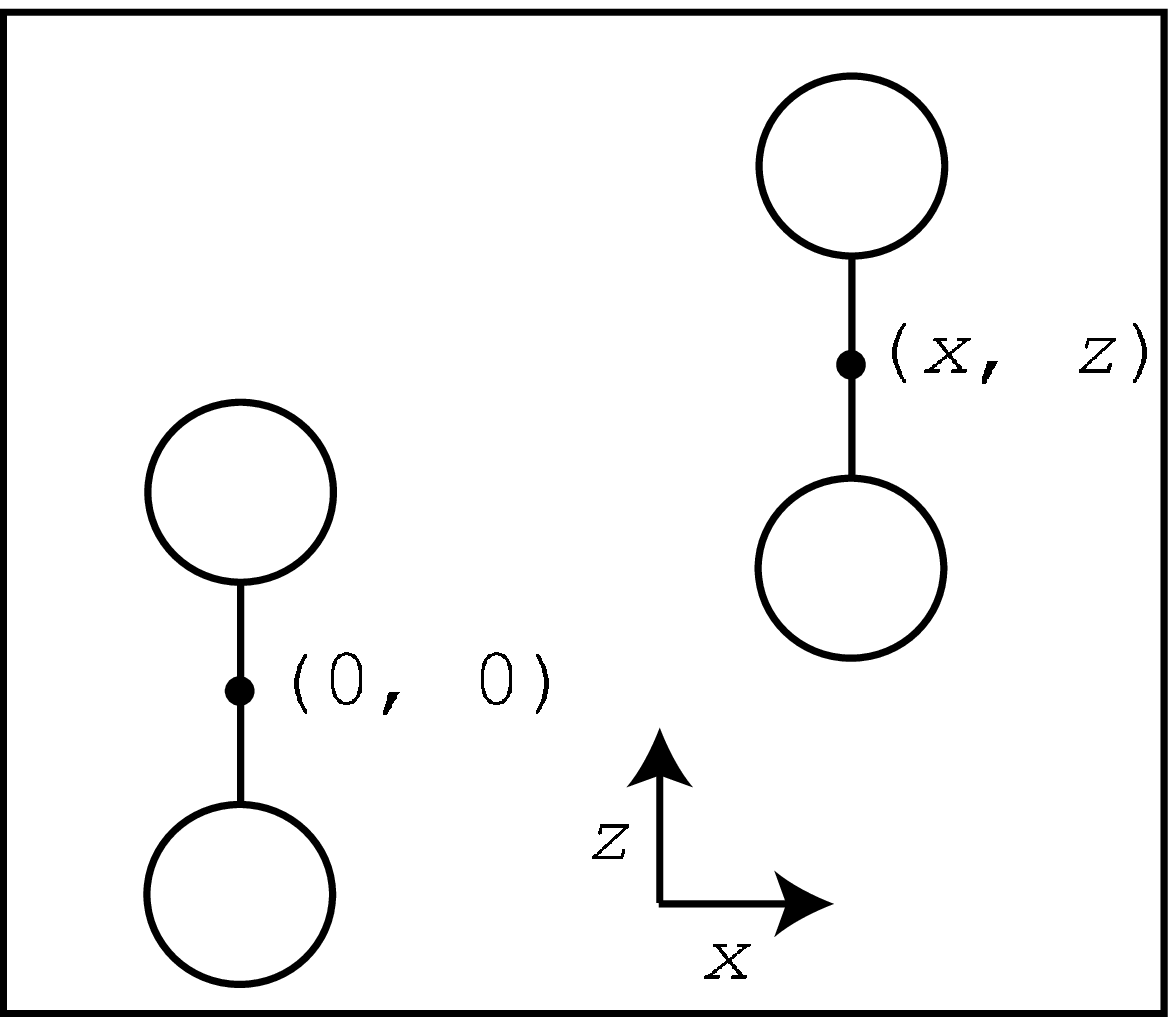}
\caption{}
\end{figure}
\begin{figure}[p]
\includegraphics[width=8cm]{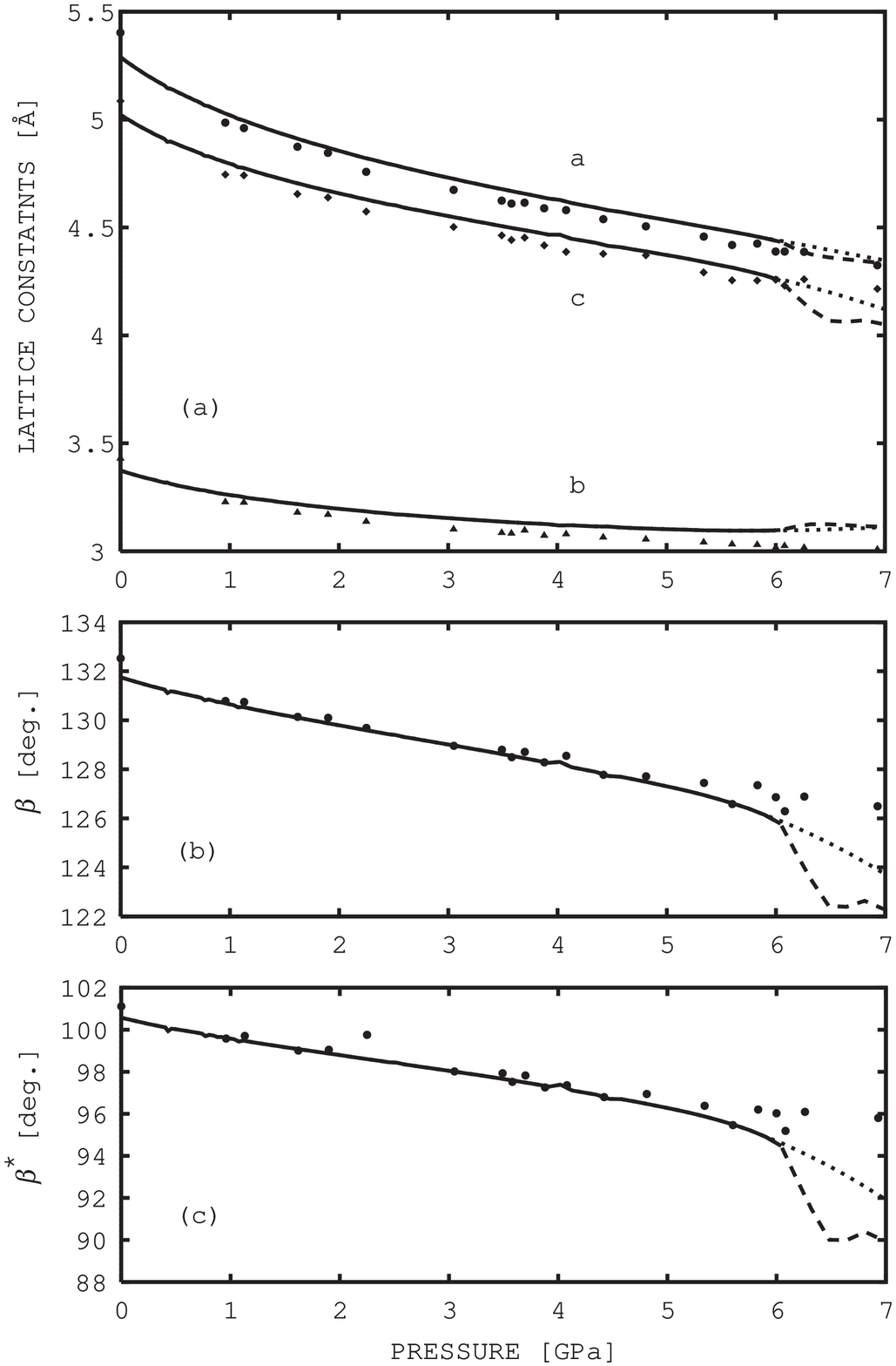}
\caption{}
\end{figure}
\pagestyle{empty}
\begin{figure}[p]
\includegraphics[width=8cm]{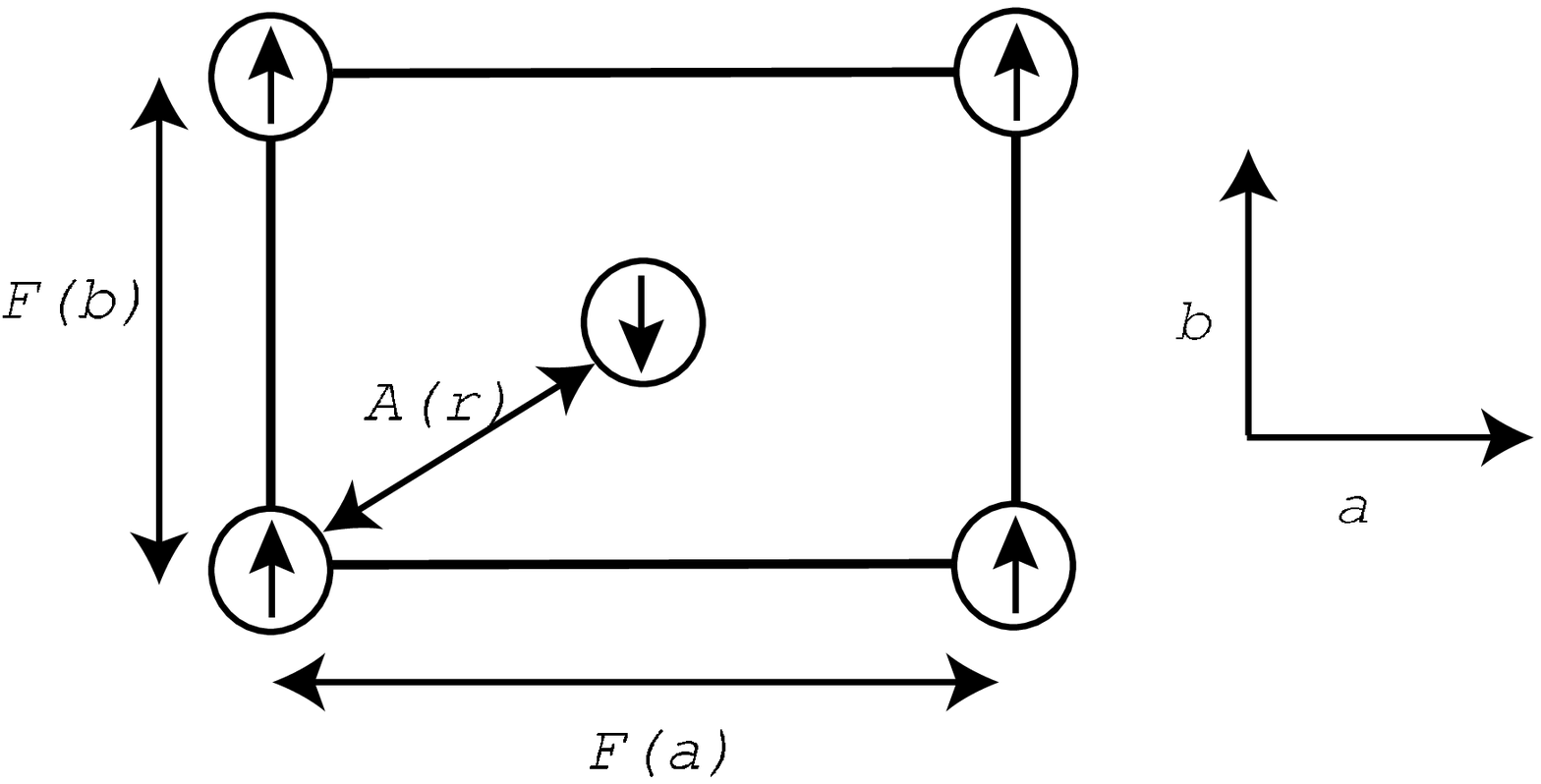}
\caption{}
\end{figure}
\begin{figure}[p]
\includegraphics[width=8cm]{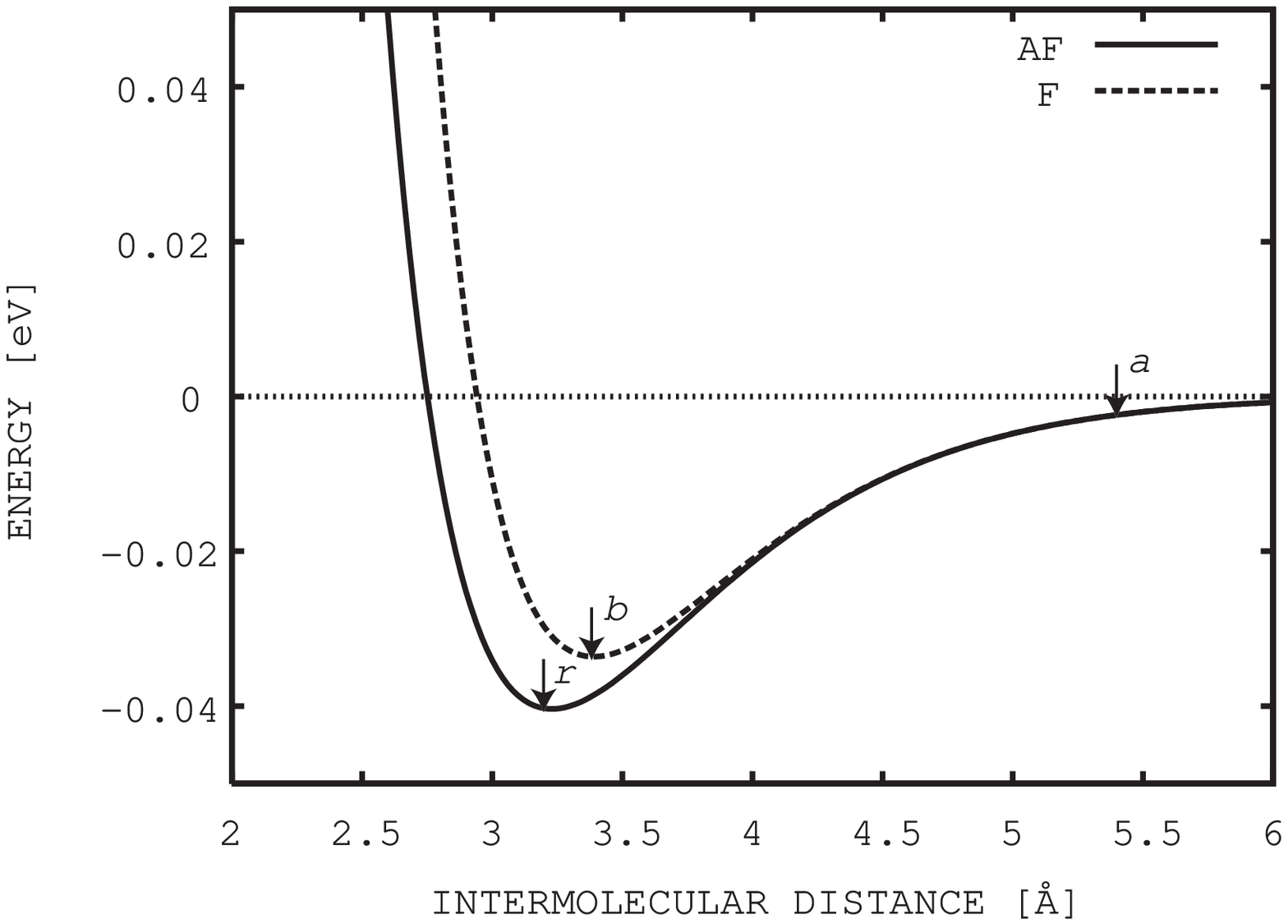}
\caption{}
\end{figure}
\begin{figure}[p]
\includegraphics[width=8cm]{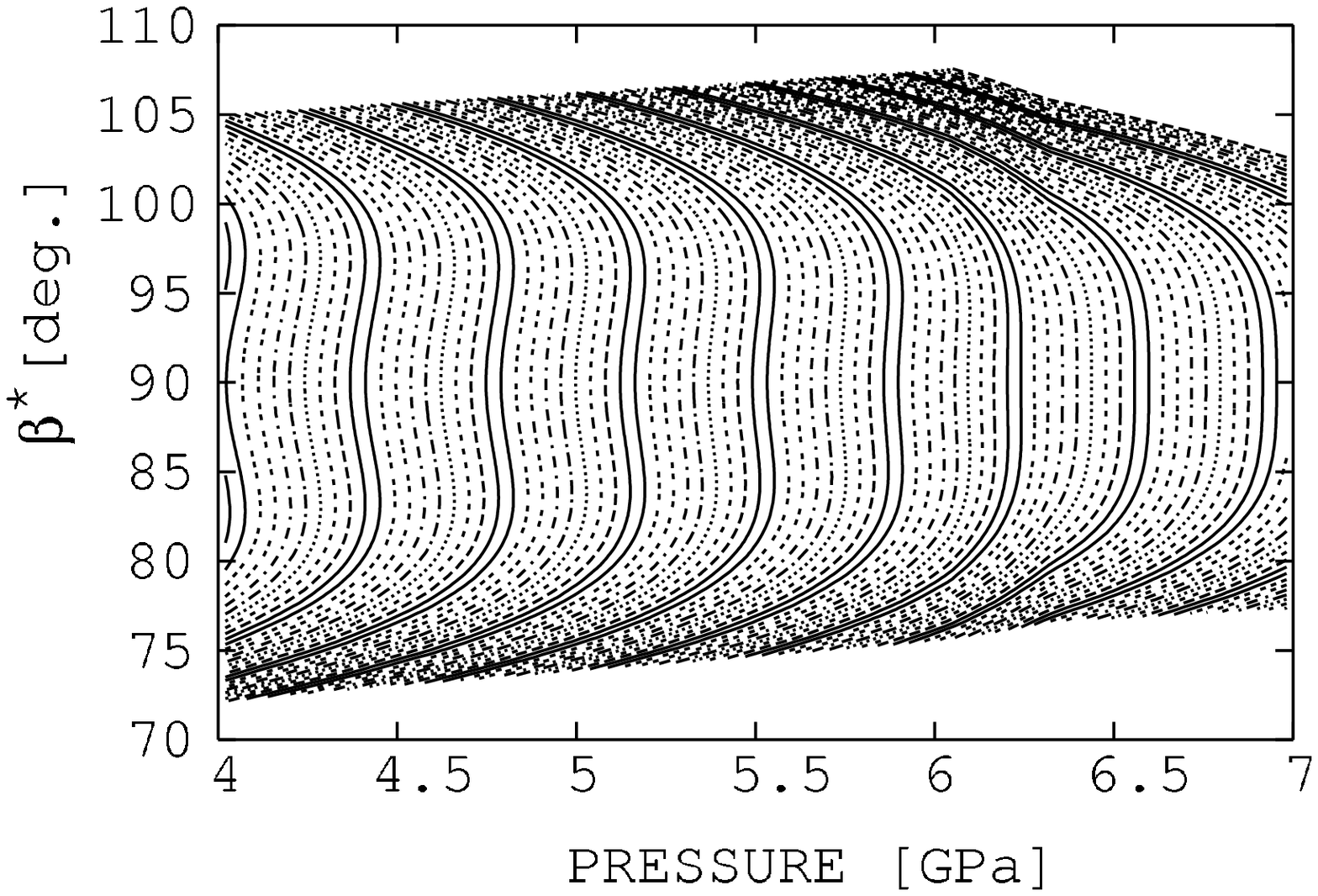}
\caption{}
\end{figure}
\end{document}